\documentclass[11pt,twoside,letterpaper]{article} 

\usepackage{times,fancyhdr}

\usepackage[dvips]{graphicx}

\usepackage{amssymb}

\usepackage{amsmath}

\usepackage{amsfonts}

\usepackage{color}

\newcommand{\PP}{\mathbf{P}}

\newcommand{\id}{\mathbf{1}}
\newcommand{\bra}[1]{\langle #1 |}
\newcommand{\ket}[1]{| #1 \rangle}
\newcommand{\braket}[2]{\langle #1 |#2\rangle}


\makeatletter 

\def\cleardoublepage{\clearpage\if@twoside \ifodd\c@page\else%

    \hbox{}%

    \thispagestyle{empty}%

    \newpage%

    \if@twocolumn\hbox{}\newpage\fi\fi\fi} 

\makeatother

\def\figurename{Figure}

\makeatletter

\renewcommand{\fnum@figure}[1]{\figurename~\thefigure.}

\makeatother

\def\tablename{Table}

\makeatletter

\renewcommand{\fnum@table}[1]{\tablename~\thetable.}

\makeatother

\begin{document}

\title{
{\begin{flushleft}
\vskip 0.45in
{\normalsize\bfseries\textit{Chapter~1}}
\end{flushleft}
\vskip 0.45in
\bfseries\scshape Understanding the Schr\"odinger equation as a kinematic statement: A probability-first approach to quantum}}

\author{\bfseries\itshape James Daniel Whitfield \thanks{E-mail address: james.d.whitfield@dartmouth.edu}\\
Department of Physics and Astronomy\\
Dartmouth College\\
Hanover, NH, US}

\date{}

\maketitle

\thispagestyle{empty}

\setcounter{page}{1}


\thispagestyle{fancy}

\fancyhead{}

\fancyhead[L]{In: Book Title \\ 
Editor: Editor Name, pp. {\thepage-\pageref{lastpage-01}}} 

\fancyhead[R]{ISBN 0000000000  \\
\copyright~2019 Nova Science Publishers, Inc.}

\fancyfoot{}

\renewcommand{\headrulewidth}{0pt}


%





\pagestyle{fancy}

\fancyhead{}

\fancyhead[EC]{J. D. Whitfield}

\fancyhead[EL,OR]{\thepage}

\fancyhead[OC]{Understanding the Schr\"odinger equation as a kinematic statement}
\fancyhead[EC]{A probability-first approach to quantum}

\fancyfoot{}

\renewcommand\headrulewidth{0.5pt} 



\begin{abstract}
Quantum technology is seeing a remarkable explosion in interest due to a wave of successful commercial technology.  As a wider array of engineers and scientists are needed,
it is time we rethink quantum educational paradigms.  Current approaches often start from classical physics, linear algebra, or differential equations. This chapter advocates for 
beginning with probability theory.  In the approach outlined in this chapter, there is less in the way of explicit axioms of quantum mechanics.  Instead the historically problematic 
measurement axiom is inherited from probability theory where many philosophical debates remain. Although not a typical route in introductory material, this route is nonetheless a 
standard vantage on quantum mechanics.  This chapter outlines an elementary route to arrive at the Schr\"odinger equation by considering allowable transformations of quantum 
probability functions (density matrices).  The central tenet of this chapter is that probability theory provides the best conceptual and mathematical foundations for introducing the 
quantum sciences.
\end{abstract}
\section{Introduction}

This chapter is meant to give a new pedagogical paradigm for understanding quantum mechanics as an extension of probability theory. Viewing quantum this way is not at all new, but it is rarely put as the central tenet of introductions to quantum theory.  Instead, quantum theory is usually introduced as an extension of classical mechanics as is done in some of the other chapters of this book. The purpose of this chapter is to provide an alternative to traditional methods for understanding the Schrodinger equation starting from probability rather than classical mechanics.  This chapter will outline the approach and its advantages when explaining the Schr\"odinger equation.  

The approach taken here does not appeal to historical derivations of the Schr\"odinger equation nor has it been tied to classical mechanics (e.g. Hamiltonian methods and Poisson brackets).  Deriving the Schr\"odinger equation this way avoids discussing the physical interpretation of the Hamiltonian and, more generally, of energy.  I put this forward as an advantage to this approach.  Energy is a difficult concept for the uninitiated and deserves its own full-fledged discussion which can be postponed until after the introduction of quantum theory depending on the composition of the audience.  By starting with a wide variety of examples from probability,  introductions to quantum theory made in physics, engineering, philosophy, mathematics or computer science from a common starting point.  
With the proliferation of quantum technology, not all students of quantum theory are physicists and may be easily confused by the energy concept. 

Moreover, in this approach, there are very few postulates explicitly required for quantum theory.  For instance, there are no explicit postulates governing evolution nor measurement.  The former is a result of kinematic constraints on the set of valid states.  The latter is implicitly postulated by choices made within probability theory.  This way quantum measurement, its interpretation, its consequences upon realization are all imported directly from probability theory.

Kinematics is the study of the motion of objects without reference to the cause of that motion.  When teaching introductory mechanics, before momentum, before energy, and even before forces, modern pedagogy begins with kinematics.   This idea can equally be applied to probability theory.  The kinematic understanding of probability theory can be nearly directly bootstrapped to quantum theory. This chapter is an exposition advocating for this approach to be adopted for training the next wave of scientists and engineers on the essentials of quantum theory.

After thoughtful consideration, I believe serious practitioners of quantum theory will recognize all the arguments found here and will likely not have any major disagreement.  However, the immediate introduction of the quantum density matrix is new to introductory pedagogy. Due to the 
newness, exercises and elementary textbook references are not directly available and would need to be adapted to this method of presenting quantum.
An effort has been made to define all terms for the beginner, but the emphasis of the chapter is for instructors and mentors who need to teach quantum to newcomers.

\section{Probability theory} 
The journey toward the Schr\"odinger equation starts not with linear algebra nor calculus but with probability theory.
Probability theory can be understood and presented at a variety of levels depending on the target audience and skills or interests of the student.  This section and the next are included in the chapter to provide the logical basis for quantum theory.  The sections on probability theory are primarily to establish notation and correct understanding of quantum theory.  However, in principle, they could be streamlined if the audience possesses sufficient background in mathematics.

A key asset of the probability-first approach to quantum is that measurement can be introduced before discussing quantum theory.  While this does not resolve nor curtail discussion of quantum foundations, it makes clear that the majority of these issues and mysteries of measurement are largely within the domain of probability theory.  This allows learners to focus on what is new within the quantum extension of probability without being confused or misled concerning the philosophy of quantum mechanics. Additionally, it allows quantum theory to be introduced without a separate ``measurement axiom."

Only the basics of probability theory are needed to arrive at a conceptual understanding of the Schr\"odinger equation.  Depending on the application areas of quantum theory planned, more or less time may be dedicated to probability theory and examples that can be revisited in the quantum domain.  For a shorter, more intuitive discussion of probability theory may allow linear algebra to be introduced and developed.  However, a more formal introduction is appropriate for practitioners interested in reading and contributing to the mathematical areas of quantum research.  

Regardless of the depth of the discussion of probability, certain concepts should be introduced at this level given the intuition that can be exposed using probability theory.  This helps give more concrete examples from probability theory when discussing mathematics.  The choice of examples can be concerted with the examples that will be introduced later in quantum theory. 

To extend probability theory to quantum theory, the notion of an $N$ dimensional orthonormal vector space basis needs to be introduced.  Then each of $N$ elementary events can be associated with the $N$ basis vectors. 
Introducing a probability vector easily allows discussing \emph{vector spaces, norms,} and \emph{inner products}.  At minimum, the notation used for these concepts should be fixed at this point. 

The choice of notation in elementary texts is split between Dirac notation and more standard mathematical notation.  Those coming from a mathematical background may be confused by some of the choices of Dirac notation.  On one hand, since its use is nearly exclusively within quantum theory,  Dirac notation requires a longer discussion regardless of mathematical preparation.  On the other hand, introducing Dirac notation immediately, allows learners to access more of the modern literature faster.  Further, the early introduction at the level of probability theory allows for many exercises before moving to quantum theory.

In this article, we utilize Dirac notation.  Briefly, in Dirac notation we denote vectors as $\vec x = \ket{x}$ and the conjugate transpose of a vector as $\bra x = ({\vec x}^*)^T={\vec x\;}^\dag$. The inner product most students have been exposed to is the dot product of vectors: 
 $\vec x.\vec y=(\vec{x})^\dag \vec{y}=\sum_j x_j^* y_j=(\vec y.\vec x)^*$. In terms of Dirac notation, one  writes $\vec x.\vec y={\braket{x}{y}}$.  

If we want $\bra{y}Ax\rangle$ to be equal to the inner product of some operator $B$ acting in the dual space such that $\bra{By}x\rangle=\bra{y}Ax\rangle$, then $B$ is called the adjoint of $A$.  In standard notation, the adjoint of $A$ is denoted $A^\dag$.  In the finite spaces we are discussing here, $A^\dag$ is the conjugate transpose of matrix $A$.

Even with more advanced audiences, it remains a good idea to explicitly define the conjugate of a complex number as $z^*=(a+bi)^*=(a-bi)$ with $i=\sqrt{-1}$ and the adjoint of a matrix as $(A^\dag)_{mn}=A_{nm}^*$. In mathematical literature, this notation for adjoint and conjugate operators is often reversed.

The pedagogical development towards quantum theory begins with introducing probability distributions as vectors. Suppose we have an experiment whose outcomes depend on chance.  
The sample space of the experiment, $\Omega$, is the set of all possible outcomes. In this chapter, we consider sample spaces with $N$ discrete exclusive outcomes. Depending on the emphasis and purposes of an introduction to quantum theory, this sample space may be considered continuous or discrete (finite or countably infinite). \footnote{For a course discussing common examples such as the hydrogen atom and real-space wave functions, the continuous formulation may be introduced here. Then the probability distribution is given by a function $p(x)$ where $\Pr(X=x)=p(x)$.  The normalization is also unity over the sample space; however, this is now defined by an integral: $\int_\Omega p(x) dx =1$.  Similarly, $\int_a^b p(x) dx = \Pr( a\leq X \leq b)$ where $a,b \in \Omega$. The $L_p$ norms are defined as $|f|_p=(\int |f(x)|^p dx))^{1/p}$.}   
Each of the $N$ elementary events is associated with orthonormal vectors $\{ \ket{e_j}\}_{j=1}^N$.\footnote{Orthonormality requires that the set is both mutually orthogonal (i.e. $\bra{e_j}e_k\rangle=0$ if $j\neq k$) and individually normalized under the $L_2$ norm induced by the inner product (i.e. $\bra{e_j} e_j\rangle=1$). The orthogonality corresponds to the exclusivity of the elementary event such that if event $j$ occurs, then event $k$ did not for all $k\neq j$.} 
Then, one can write a probability vector as
\begin{equation}
    \ket{p}=\sum_j p_j \ket{e_j}
\end{equation}
The numerical components of the probability vector are given by $p_j=\braket{e_j}{p}$.  For use within probability theory, each $p_j$ satisfies $0\leq p_j\leq 1$ and satisfy the normalization condition $\sum_j p_j=1$.  Its important to highlight that $\{\ket{e_j}\}$ is fixed but otherwise arbitrary.  

\subsection{Measurement} 
Obtaining the outcome of an experiment is a \emph{measurement}.  The actual act of measuring the outcome of an experimental realization is not necessary for introducing quantum theory.  By maintaining a conceptual link to probability theory when approaching quantum theory, connecting experiments and outcomes remains an exercise in probability theory.  Then questions such as assigning, updating, and measuring a probability distribution before, after or even during a measurement are the same in quantum theory.  We make the argument clearer when discussing coherence and decoherence of quantum states.  

In standard treatments of quantum theory, measurement is often tacked on as a postulate of quantum theory.  However, with quantum theory as an extension of probability theory the measurement postulate is obtained from ordinary probability theory.  By examining the quantum kinematic constraints, we will obtain the Schr\"odinger equation.  Before moving on to quantum generalizations, the discussion of kinematic constraints can be prepared by examining the kinematics of probability distribution functions.

With the definitions and requirements of probability distributions, we can consider the possible transformations and evolutions of probability distributions without reference to the causes of these changes. It is this agnosticism that will help expose a broad understanding of quantum theory but may introduce difficulties in connecting to applications.  This can be ameliorated by choosing examples and illustrations appropriate for the target audience.

\section{Kinematics of probability distributions}\label{sec:pkinematics}

Schr\"odinger's equation and all other quantum equations of motion must obey kinematic constraints that ensure the form of the quantum state remains valid.  This kinematic approach is first illustrated using probability vectors in this section, which is later generalized to quantum theory in \ref{sec:qkinematics}.

In this section, we wish to consider transformations that take one valid probability distribution $\ket p$ to another valid probability distribution $\ket{p'}$.  We consider several classes of transformations that provide direct parallels to the quantum kinematic discussion.  The change of basis concept is essential to the development of the Schr\"odinger equation as a kinematic expression. Also included in this section is a discussion of the master equation describing differential transformations of probability distributions.  The quantum extension of this differential equation goes beyond the Schr\"odinger equation and, thus, is outside the scope of this chapter.  However, in the context of mentoring or teaching, this provides a nice way to open the discussion of stochastic quantum processes and general evolution equations for quantum systems in contact with the environment.

The argument of this chapter requires viewing the change of basis formula for probability vectors as just an exchange of labels.  The arbitrary map from elementary events to basis vectors allows us to rearrange the vector without changing any of the values by relabelling the event or relabelling the basis vectors. This is done using a permutation $\PP \in S_m$ where $S_m$ is the symmetric group of permutations of $m$ objects.  

If we consider each permutation as an $m\times m$ matrix,  \footnote{For example, a representation of $S_3$ is given by $\left\{\mathbf{P}_{12}=\begin{pmatrix} 0 & 1 &0\\1&0&0\\0&0&1\end{pmatrix},\mathbf{P}_{23}=\begin{pmatrix} 1 & 0 &0\\0&0&1\\0&1&0\end{pmatrix}\right\}$ and the matrix products $\PP _{23} \PP _{23}$, $\PP _{23}\PP _{12}$, $\PP _{12}\PP _{23}$, $\PP _{12}\PP _{23} \PP_{12}$.} then it is easy to introduce matrix multiplication exercises using Dirac notation.  Simple manipulations such as 
\begin{eqnarray}
p'_j&=&\sum_k \PP _{jk}p_k\\
\langle e_j | p'\rangle&=&\langle e_j | \left(\PP \ket p\right)\\
&=&\sum_k \langle e_j| \PP\ket{e_k}\langle e_k \ket p 
\end{eqnarray}
are useful for more elementary audiences.  Here, it is also useful to introduce the resolution of the identity $\mathbf{1}=\sum_j \ket{e_j} \bra{e_j}$ and emphasize that it holds whenever $\{e_j\}$ is an orthonormal basis.  This can help simplify later discussion and clarify notations for more advanced students unfamiliar with Dirac notation.

Stochastic matrices are introduced via convex combinations of permutation matrices in this section.  This is not the only way to introduce transformations that preserve the kinematics constraints of probability vector transformations.  Moreover, we do not reach the full class of stochastic matrices but rather bistochastic transformations.  For the purposes of reach the Schr\"odinger equation, it is not necessary to go beyond bistochastic matrices.  

To get a transformation that preserves the validity of probability vectors, suppose with probability $w_j$ permutation $\PP_j$ is applied ($\sum_j w_j=1$).  Then the transformation
\begin{equation}
    M=\sum_j w_j\PP_j
\end{equation} 
can act on $\ket p$ giving $M\ket p=\ket{ p'}$ as a valid probability distribution.  Matrices such as $M$ that transforms valid probability distributions to valid probability distributions are called called \emph{stochastic matrices}.\footnote{This construction of stochastic matrix $M$ as a convex combination of permutation matrices is a way to characterize all bistochastic matrices \cite{Horn}. More general one-sided stochastic matrices are possible and are the reason that $\sum_j M_{ji}$ is not necessarily unity in \eqref{eq:master}.  Many probability books such as \cite{Snell} prefer the stochastic matrices written such that $\tilde{M}_{ij}=\Pr(i\rightarrow j)$.  In this case, probability vectors are row vectors $\mathbf{x}$ multiplying from the left $\mathbf{x} \tilde{M}$ rather than column vectors multiplying from the right $M\ket p$. The bistochastic matrices can work with probability vectors as column vectors on the right or row vectors on the left; hence the prefix `bi-'.  Here, we only consider probability vectors as column vectors. } 
The use of the values $M_{ij}$ as transition probabilities is a useful point of reference before moving to quantum theory 
\begin{equation}
    M_{ij}=\Pr(j\rightarrow i) = \Pr(X'=i|X=j)=\frac{\Pr(X'=i\text{ and }X=j) }{\Pr(X=j)}
\end{equation}
This serves as a way to introduce conditional probabilities $P(A|B)$ and give a wealth of examples to illustrate matrix multiplication in a more concrete fashion with transition matrices.

Since the Schr\"odinger equation determines differential changes,
it is useful to describe valid transforms of probability distributions via a differential equation.  The typical way to characterize the infinitesimal changes of a probability density function is in terms of a master equation. In such equations, the rate of change in one component of the probability vector is the rate of probability entering that component minus the rate of probability decreasing from that component. With the rate of going from state $i$ to state $j$ as $M_{ij}$, then in matrix form, we write
\begin{align}
\begin{array}{c}\textrm{component}\\
\text{change}
\end{array}
 &= [ \text{rate in}] - [\text{rate out}]\\
dp_k&= \left[ \sum_j M_{ij}p_j\right]  - \left[\left(\sum_jM_{ji}\right)p_i\right]
    \label{eq:master}
\end{align}
We can further summarize this statement by first defining the degree matrix as $D_{kk}=\sum_jM_{jk}$ and then the generator of the evolution is given by $L=M-D$ such that
\begin{equation}
    L\ket p =\frac{d\ket p}{dt}
    \label{eq:Wgen}
\end{equation}
The real variable $t$ is used to parameterize the evolution under $L$.  For simplicity, we assume that the matrix $L$ is time-independent.

We can integrate the equation following
\begin{eqnarray}
L\ket{p}=\frac{d\ket{p}}{dt} &\implies & Ldt=\frac{d\ket{p}}{\ket{p}}\\
&\implies&Lt=\log \ket{ p_t} - \log \ket{p_0}\\
&\implies& e^{Lt}=\exp(\log \ket{p_t} -\log \ket{p_0})\\
&\implies &e^{Lt}\ket{p_0} = \ket {p_t}
\label{eq:evol}
\end{eqnarray}
For all positive real values of $t$, $M(t)=\exp(Lt)$, is a stochastic matrix.  The analog of  \eqref{eq:evol} in quantum theory is beyond the Schr\"odinger equation; however, its integration is instructive.  The formulation of \eqref{eq:Wgen} has much more in common with the Schr\"odinger equation than the standard equations found in a first course on classical mechanics.  From a pedagogical perspective, the derivation also offers an opportunity to introduce more mathematical concepts including integration and the matrix exponential.  

If readers are unfamiliar with linear algbera, it might be useful to introduce matrix mechanics using a programming environment (e.g. Python).  The numerical tools built or utilized here should be planned to have their quantum extensions discussed.  This allows students to test and play with the concepts and equations on small numerical examples before reaching quantum theory.  The eigendecomposition, Taylor expansion, and matrix functions can be introduced alongside \eqref{eq:evol}.

\section{Quantum theory via probability}\label{sec:qtheory}

With the sufficient notions and definitions from probability theory and linear algebra established, we can proceed to discussing quantum theory as a generalization of probability theory. The quantum probability density matrix in quantum mechanics is the direct generalization of the probability distribution vector. 

\begin{figure}
    \centering
    \includegraphics[height=10em]{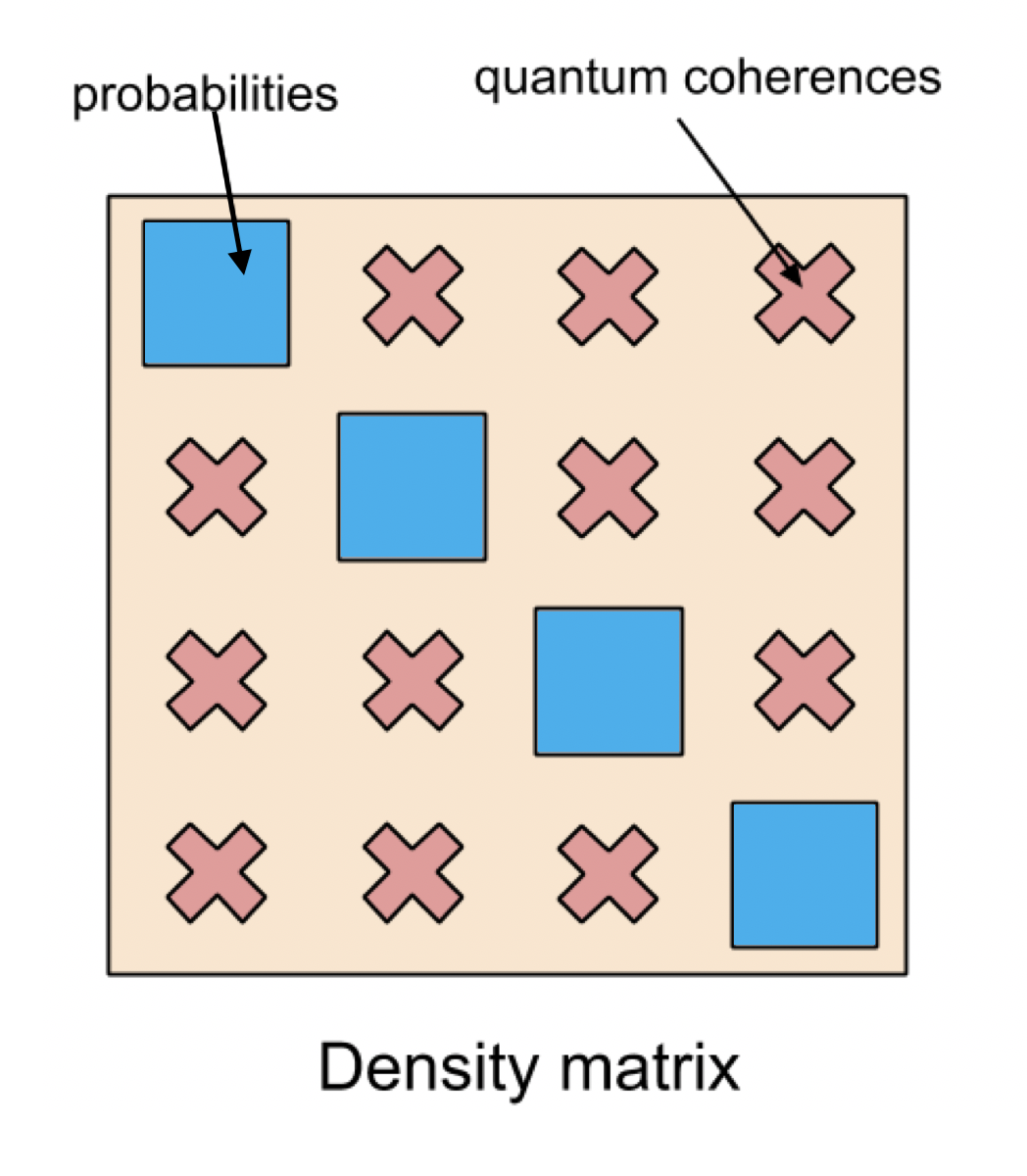}
    \caption{A graphical illustration of the density matrix.  The diagonal elements form a valid probability distribution and give the propensity for an outcome to be realized upon measurement in the depicted basis. The off-diagonal element are called coherences.  This illustration is particularly useful for elementary discussions with less mathematical involvement.}
    \label{fig:rho}
\end{figure}

Quantum probability distributions, often referred to as the density operator or density matrix, are commonly denoted with $\rho$.  Valid density matrices must satisfy the normalization condition $\sum_i \rho_{ii} = 1$ which replaces the $L_1$ condition of probability vectors.  The probability vector consists of only positive numbers.  The comparable statement for density matrices is that for every normalized (possibly complex) vector $\vec v$, the expectation value $\langle v |\rho | v\rangle$ is real and greater or equal to zero (see Fig. \ref{fig:rho}).  
\footnote{Alternatively, this can be stated more succinctly: the density matrix is Hermitian, positive semi-definite, and has a normalized trace.} 

Some introductory material on quantum theory suggests that the wave function is analogous to the probability vector with the key difference being the $L_2$ norm is used for the former and the $L_1$ norm for the latter.  In the probability first approach, the $L_1$ norm remains applicable to the diagonal of the density matrix.  The off-diagonal elements are called coherences and are strictly quantum.  The development of coherences is what allows for a separation between quantum and ordinary probability theory.  

\begin{figure}
    \centering
    \includegraphics[height=10em]{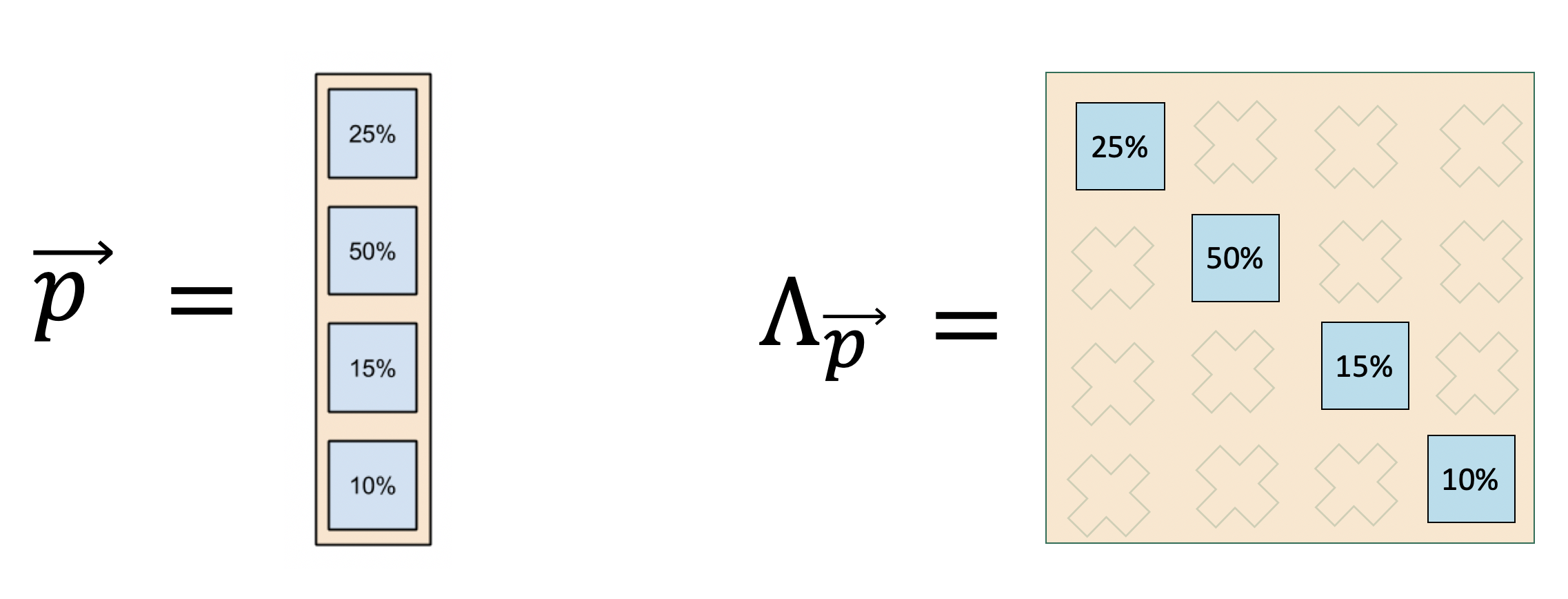}
    \caption{Here, a valid probability vector and the corresponding quantum density matrix are given.  Note that the underlying basis for both the matrix and the vector are the same.}
    \label{fig:qlift}
\end{figure}

The density matrices formed by taking a probability vector and converting to a matrix with the same probability values along the diagonal (i.e. $\rho=\Lambda_{\vec p}=\textrm{diag}(\vec p)$) with zeros elsewhere. See Fig. \ref{fig:qlift} for an example of the correspondence.  The set of matrices formed by taking probability vectors and placing their entries along the diagonal are all valid density matrices. 

Due to the conditions on valid density matrices, there always exists an orthonormal basis where the density matrix is diagonal. The eigenbasis of the density matrix is  called the \emph{natural} basis of the quantum system since it is defined without external references.  In the natural basis, say $\{\psi_j\}$, there are no coherences and the density matrix can be written
\begin{equation}
    \rho=\sum p_j \ket{\psi_j}\bra{\psi_j}
    \label{eq:qprob}
\end{equation}
with $\sum p_j=1$. The basis of the density matrix in Fig. \ref{fig:qlift} is the natural eigenbasis and the diagonal elements give the probability for obtaining a particular basis vector upon measurement.

If the reader notices, wave functions have not been mentioned explicitly.  For audiences with some exposure to quantum theory, it is useful to connect wave functions back to the density matrix alongside their introduction.  Then the density matrix in its eigenbasis can be described as a probability distribution over wave functions.  The apt illustration of \eqref{eq:qprob} is a wave function generator that outputs state $\psi_j$ with probability $p_j$.

Another difference between a quantum density matrix and a probability vector are the allowable changes to the basis.  Previously, we allowed a change of basis defined by a permutation matrix $\PP$. In quantum theory, we can now rotate the basis without changing the eigenvalues found in \eqref{eq:qprob}.  

Thus, quantum theory is introduced using a quantum probability density operator that generalizes the probability distrubution vectors.  We can now turn toward generalizing the possible change of basis allowed under quantum theory. This will lead us directly to the Schr\"odinger equation.

\section{ Transformations of the quantum probability density distributions}\label{sec:qkinematics}
In this section, the allowable transformations of quantum density matrices is approached in the same fashion as in section \ref{sec:pkinematics}   We first illustrate the generalization of probabilistic transforms in the quantum density matrix setting.  Then the generalization of the change of basis is introduced. Finally, the allowable transformations of quantum density matrices are reached in direct parallel to the probability case.

This development is logical and sensible for introducing quantum theory kinematically. However, the generalization of stochastic matrices to the quantum domain is more than is necessary for the Schr\"odinger equation.  For the Schr\"odinger equation, it is enough to appreciate the continuous change of basis allowable for density matrices without composition.

Earlier, permutations served as the change of basis for probability vectors where $\PP\vec p=\vec p\,'$.  To effect the same transformation on diagonal density matrix, $\Lambda_{\vec p}$, we use
\begin{equation}
    \PP \Lambda_{\vec p} \PP^\dag = \Lambda_{\vec p\,'}
    \label{eq:perm_action}
\end{equation}
Similarly, we have for $M=\sum w_k \PP_k$ the equivalent in terms of quantum density operators.
\begin{equation}
    \sum_k w_k\PP_k \Lambda_{\vec p} \PP_k^\dag = \Lambda_{M\vec p}
    \label{eq:cl_channel}
\end{equation}

Now instead of just permuting the basis vectors, we may additionally rotate the basis vectors from set $\{e_j\}$ of basis vectors to another $\{f_j\}$.  At this point, many examples can be used to illustrate the notion of differing basis used to describe the same object. 

For a chemical audience, atomic orbital basis and the molecular orbital basis provide an example of two bases for describing a single density matrix. In solid state physics, the change of basis from $k$-space to the real space is an instructive use of the Fourier transform.  Examples in the lab frame and in the rotating frame are appropriate if magnetic resonance is planned as part of the quantum course.  Two-level systems (i.e. qubits) provide a wealth of illustration, examples and demonstrations.  The Bloch sphere and the Stern-Gerlach experiment are two examples that also help visualize the idea of multiple bases describing the same object. A simple example is found using plane rotations in two dimensions which provide mathematical and graphical illustrations and exercises. 

Regardless of the examples chosen, the change of basis is represented by a matrix that converts each orthonormal basis vector from one set $\{e_j \}_{j=1}^N$ to another orthonormal set $\{f_j\}_{j=1}^N$.  This is arranged in matrix form as
\begin{equation}
    U=\sum_j \ket{e_j}\bra{f_j}=\sum_j (\vec{f}_j) (\vec{e}_j)^\dag
\label{eq:U_cob}
\end{equation}
In general, we know that $U$ is a valid transformation of the basis if $U^\dag=U^{-1}$. Such matrices are called \emph{unitary} and pervade quantum theory.

The change of basis is required in many settings.  For example, the quantum state may have been prepared in basis $\{e_j\}$ but the experiment is performed in basis $\{f_j\}$.  Then the probabilities determining the outcome of experiments is then given by the diagonal of $U\Lambda_p U^\dag$ with $U$ given in \eqref{eq:U_cob}.  

Similar to the characterization of stochastic matrices, we can characterize more general transformations from one valid density matrix $\rho$ to another $\rho'$ by changing the initial basis with various probabilities. 
\begin{equation}
    \rho'=\sum_k w_k U_k\rho U_k^\dag 
    \label{eq:qm_channel}
\end{equation}
This is the quantum generalization of equation \eqref{eq:cl_channel}.\footnote{Similar to the footnote following \eqref{eq:cl_channel}, the expression in \eqref{eq:qm_channel} is also not the most general.  The broadest characterization of transformations from $\rho$ to $\rho'$ is characterized by the Kraus representation: $\sum_k E_k \rho E_k^\dag$ where the only constraint on the operators $\{E_k\}$ is that $\id=\sum E_k^\dag E_k$.  This is clearly satisfied for $\{E_k=\sqrt{w_k}U_k\}$ as found in \eqref{eq:qm_channel}.} 

The arrival of equation \eqref{eq:qm_channel} is quite beautiful in that it structurally echoes the probability evolution in \eqref{eq:cl_channel}.   Noting that permutation matrices are also unitary matrices, the \eqref{eq:cl_channel} reduces to a special case of \eqref{eq:qm_channel}.
Here many asides could be made depending on the audience and its aims.  Noise, thermodynamics, entanglement, communication, and many other topics can be introduced once \eqref{eq:qm_channel} has been introduced.  

\begin{figure}
    \centering
    \includegraphics[width=\textwidth]{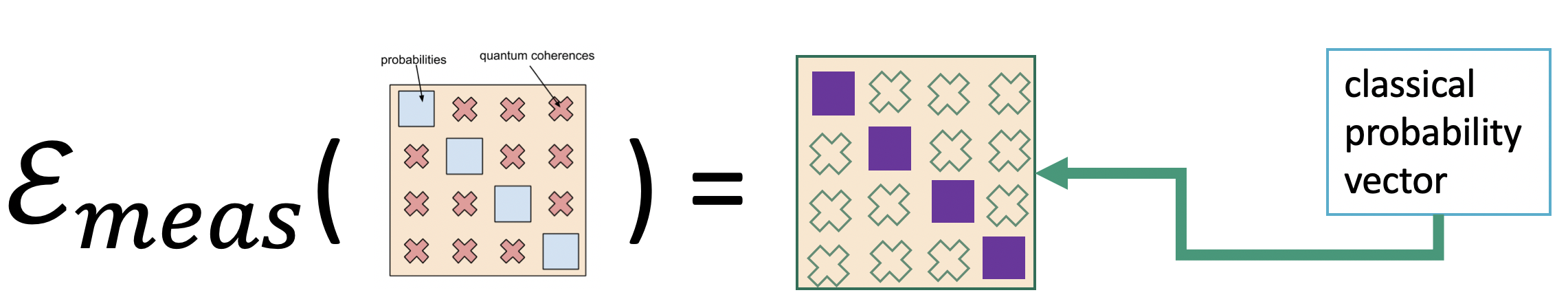}
    \caption{A graphical illustration of the quantum aspects of measurement. The quantum channel $\mathcal{E}_{meas}$ transforms a density matrix to a probability distribution in the basis of the measurement.  Then the outcome realization and its consequences are issues strictly within the domain of ordinary probability theory. }
    \label{fig:meas}
\end{figure}

A return to the discussion of measurement is also appropriate following \eqref{eq:qm_channel}.  It is sufficient to consider measurements as those of ordinary probability theory using the diagonal elements of the density matrix in the right basis.  Decoherence theory is also an instructive tangent where decoherence can be summarized as the loss of the coherences in the density matrix. With all the coherences zero, the density matrix is an ordinary probability distribution.   A measurement channel is illustrated in (Fig. \ref{fig:meas}) but its usefulness in elementary courses can obscure the connection to reality without sufficient examples.  Connecting the mathematical formalism to physical examples is an important exercise that can occasionally confound newcomers.

In many ways, a natural direction continues with the kinematic derivation of the quantum analog of \eqref{eq:master}.  Then the Schrodinger equation appears as the non-dissipative portion of the evolution.  This is a choice that should be balanced with the eventual goals of learning quantum theory.  The lengthier discussion does not add much, if it is unlikely to be used later in the course or in the research trajectory of the students.

With both the concept of the density matrix and its transforms as extensions of probability theory, we now turn to obtaining the Schr\"odinger equation as kinematic statement.

\section{The Schr\"odinger equation}
The structure of stochastic and quantum transformations are highlighted by the parallels of \eqref{eq:cl_channel} and \eqref{eq:qm_channel}.  The quantum master equation that generalizes \eqref{eq:master} is unnecessary for the purposes of understanding the Schr\"odinger equation.  It is enough to consider the special case of \eqref{eq:qm_channel}.

To arrive at the Schr\"{o}dinger equation, we just need to find the infinitesimal basis rotation of a specific quantum event $\ket{\psi}$ that occurs with probability 1. 
Such a quantum state is given by $\rho=\ket{\psi}\bra{\psi}$. Next, we parameterize the change from one density matrix to another by a family of single unitary transformations $U(t)$ where $t$ represents time:
\begin{eqnarray}
    \rho_t&=&U(t) \rho U(t)^\dag\\
          &=&U(t) \ket{\psi} \bra{\psi}  U(t)^\dag
          \label{eq:18}
\end{eqnarray}
Since the direct and dual spaces are redundant, we can simplify \eqref{eq:18} as
\begin{equation}
    \ket{\psi_t}=U(t)\ket{\psi}
    \label{eq:preSE}
\end{equation}
For the final step, we will need an additional fact about unitary matrices that connects them with the matrix analog of a real number.  First, one can consider a minimal case where the unitary matrix consists of a single element. Then we can satisfy condition $uu^*=uu^\dag=u^* u=\id$ with $u=\exp(-i\theta)$ as long as $\theta$ is real. 

To motivate the matrix equivalent, we define the matrix analog of the real part of a complex number.  For arbitrary matrix $A$, the Hermitian part is $\frac12(A+A^\dag)$ in analogy to scalar equation Re$(z)$=Re$(a+bi)=\frac12 (z+z^*)=a$. For completeness, it may be worthwhile to continue the analogy by introducing skew-Hermitian matrices $\frac12(A-A^\dag)$ are the matrix analog of the imaginary part of a complex number. The property that $H=H^\dag$ is in analogy with $a=a^*$ in the scalar case. From there, the analog with $u=\exp(-i\theta)$ continues as
\begin{equation}
    U=\exp(-i H)=\sum_{n=0}^\infty \frac{\left(-i H\right)^n}{n!}
    \label{eq:U}
\end{equation}
Some care must be given when taking the matrix exponential, in that one cannot exponentiate element-wise.  This is most simply explained using eigendecompositions or alternatively using the Taylor expansions as in righthand side of the final equality of \eqref{eq:U}.

Now to obtain the analog of \eqref{eq:evol}, we still need to generate a family of unitary matrices connected via a real-valued parameter.  This is easily accomplished by extending \eqref{eq:U} with real parameter $t$
\begin{equation}
    U(t)=\exp(-i H t)
    \label{eq:Ut}
\end{equation}
Unlike the situation of stochastic evolution, this parameter can be positive or negative.  Thus, the sign convention of the last equation is arbitrary, but our convention here follows standard practices of quantum theory.

Finally, we can take the derivative of \eqref{eq:preSE} to arrive at 
\begin{eqnarray}
    \frac{d\ket{\psi_t}}{dt}&=&\frac{d}{dt}U(t)\ket{\psi}\\
    &=&-iH U(t)\ket{\psi}\\
    \frac{d\ket{\psi_t}}{dt}&=&-iH\ket{\psi_t}
    \label{eq:SE}
\end{eqnarray}
Finally, we have arrived at the time-dependent Schr\"odinger equation in \eqref{eq:SE} without appeal to classical mechanics. 

It is worthwhile to point out that the probabilistic differential equation of motion described earlier \eqref{eq:master} is not analogous to the Schr\"odinger equation.   This is because the Schr\"odinger equation describes how the rotation of the vector space basis occurs infinitesimally.  Given that there is no continuous transformation between discrete permutation matrices, there is no direct analog in probability theory.  However, there is a direct quantum analog of \eqref{eq:master} resulting in a differential form for the transformations in \eqref{eq:qm_channel} which is called the Lindblad equation.

\section{Outlook}

In the approach that this chapter advances, quantum theory is seen as an extension of probability theory. Foundational issues such as the epistemic (“knowledge”) or ontic (“being”) nature of the probability function can be addressed here before approaching quantum theory.  There are interesting foundational and philosophical issues concerning the interpretation of probability theory which can be identified and set aside before discussing quantum theory.  Depending on the tastes of a lecturer or the learners, segues into the frequentist and inference interpretations of probability theory and of quantum theory can be explored in parallel.  

Regardless of what the reader or teacher chooses to pursue next, this chapter has focused on understanding the Schr\"odinger equation from a kinematic point of view.  
Introductory mechanics is taught at the high school level whereby all necessary analytic tools are introduced as needed, and the course is even taught without calculus.  The prescription here aims to give an approach that allows a course to be taught at various levels depending on the sophistication of the student but following a basic structure similar to introductory mechanics.

Questions of prerequisites, necessary mathematical background, and whether to begin with wave mechanics or discrete systems are answered as followed: no prerequisites should be required for an introductory course beyond a solid high school education and an ambition to learn.  To be sure, linear algebra, calculus, and differential equations are important for any serious study of physics or related technical subjects, but for an introduction to quantum theory, they are superfluous and can be introduced as needed.

Depending on the nature of the course and the goals of the learner, there is a wide variety of directions to continue following a probability-first introduction to quantum theory.  Since the energy concept was not stressed in this presentation, it is an natural next step.  Moreover, given that this presentation has already introduced probability theory and the density matrix, open systems can be introduced immediately after the discussion of energy.  Thermodynamics could also follow closely behind. 

The notion of entropy and connected concepts in thermodynamics can be introduced alongside the probability introduction and re-examined in the quantum context.  There are three sensible reasons to consider entropy directly after introducing quantum theory: (i) it can be introduced at various levels directly using probability theory, (ii) historically, the idea of quantization has its origins in statistical mechanic considerations (i.e.~the ultraviolet catastrophe that led Planck to introduce quantization as a concept), and (iii) entropy opens the door to discussion of entanglement entropy and algorithms for exploiting low entanglement \cite{Veretraete}.

There may remain reluctance to switch given the overhead it places on crafting new notes, selecting new texts, and designing the correct scaffolding for various levels and backgrounds.  However, I hope the elegance alone persuades some teachers and mentors to switch to a probability-first approach.

\label{lastpage-01}

\end{document}